\documentclass{amsart}

\usepackage{pxfonts}
\usepackage[T1]{fontenc}
\usepackage[utf8]{inputenc}
\usepackage[english]{babel}
\usepackage{amsmath}
\usepackage{amsfonts}
\usepackage{amssymb}
\usepackage{ragged2e}
\usepackage{setspace}
\usepackage{graphicx}
\usepackage{tikz}
\usepackage{tabu}
\usepackage{booktabs}
\usepackage{array}
\usepackage[linesnumbered,ruled,vlined]{algorithm2e}
\usepackage{hyperref}

\newcommand{\F} {\mathbb{F}}
\newcommand{\A} {\mathbb{A}}

\renewcommand{\P} {\mathbb{P}}
\renewcommand{\O} {\mathcal{O}}
\newcommand{\Res} {\text{Res}}
\renewcommand{\div}{\text{div}}

\begin{document}

\large

\title{A New Method for Geometric Interpretation of Elliptic Curve Discrete Logarithm Problem}

\maketitle

\vspace{2mm}

\begin{center}

	Daniele Di Tullio $^1$ \& Ankan Pal $^2$ \\ 

	$^1$ Department of Mathematics, Roma Tre, Italy \\ 

	$^2$ Department of Mathematics, University of L'Aquila, Italy \\

	E-mail:	danieleditullio@hotmail.it $^1$\\
 
 			ankanpal100@gmail.com $^2$

\end{center}

\vspace{2mm}

\noindent \textbf{Abstract:} In this paper, we intend to study the geometric meaning of the discrete logarithm problem defined over an Elliptic Curve. The key idea is to reduce the Elliptic Curve Discrete Logarithm Problem (EC-DLP) into a system of equations. These equations arise from the interesection of quadric hypersurfaces in an affine space of lower dimension. In cryptography, this interpretation can be used to design attacks on EC-DLP. Presently, the best known attack algorithm having a sub-exponential time complexity is through the implementation of Summation Polynomials and Weil Descent. It is expected that the proposed geometric interpretation can result in faster reduction of the problem into a system of equations. These overdetermined system of equations are hard to solve. We have used F4 (Faugere) algorithms and got results for primes less than 500,000. Quantum Algorithms can expedite the process of solving these over-determined system of equations. In the absence of fast algorithms for computing summation polynomials, we expect that this could be an alternative. We do not claim that the proposed algorithm would be faster than Shor's algorithm for breaking EC-DLP but this interpretation could be a candidate as an alternative to the 'summation polynomial attack' in the post-quantum era.          
	
\vspace{2mm}
	

\section{Introduction}

\noindent	Elliptic curves defined over a finite field ($E({\F}_{p})$) essentially are rich mathematical structures which result in their ubiquitous use in Number Theory, Cryptography, and Algebraic Geometry. EC-DLP has myriad applications and it is used to design most of the classical crypto-systems involving Elliptic Curves. Initially, index calculus methods were used to attack the EC-DLP. The major impediment to such attacks was to find a computationally optimized factor base $\mathcal{F}$. There were considerable improvements in this direction using baby-step-giant-step method. In the case, in which the cardinality of the set of the rational points is a smooth number the Pohlig-Hellman attack \cite{W08} is quite efficient. Pollard's $\rho$ and $\lambda$ methods were another approach for designing the attacks \cite{P78}. The case for supersingular elliptic curves was solved by the use of Weil Paring (MOV Attacks) \cite{MOV93}. The case for curves of cardinality $p$ (trace = 1) is easily reduced to DLP over $\mathbb{Z}/p\mathbb{Z}$. In the other cases the EC-DLP is still hard to attack. Presently, the best method known attack algorithm having a sub-exponential time complexity for particular cases is through the implementation of Summation Polynomials and Weil Descent \cite{GS99} \cite{S04} \cite{D13}. 

\vspace{2mm}

\noindent   In 2000, approaches which use tools from Algebraic Geometry were presented for EC-DLP by Galbraith and Smart \cite{GS99}. In this paper, Weil descent was used to attack EC-DLP. The use of Weil restrictions, modified the problem of solving EC-DLP on the elliptic curve defined over a finite field extension to solving it on the jacobian of a higher genus curve but defined over a smaller base field \cite{G09}. Sub-exponential time algorithms exists for solving the EC-DLP over higher genus curves \cite{F90}. This was an approach which was successful for many particular cases. In his seminal paper of 2004, Semaev \cite{S04} constructed summation polynomials to solve EC-DLP. Grobner basis was used for solving the system of equations that arose from such problems. A working attack algorithm of sub-exponential time complexity for small dimensions was developed by Gaudry \cite{G09} using the concepts of Summation Polynomials and Weil Restrictions.

\vspace{2mm}

\noindent   Solving EC-DLP generally does not have fast algorithms. All the existing methods rely heavily on computing the summation polynomials. This reliance is avoided and a new method for attacking EC-DLP is elucidated. The major drawback of the intersection method is that it produces over-determined system of equations which have high time complexity. We have used Grobner basis and F4 (Faugre) algorithm \cite{F99} to solve these system of equation. This has not produced good results for large primes (the characteristic of the base field). For primes less than 500,000, the intersection method produces good result. We have implemented the intersection method with the help of MAGMA \cite{BC93}. In the post quantum era, we expect that we can solve the system of equations using the methods as illustrated in \cite{FHKKKP17} \cite{AD07} \cite{HHL09}. We do not expect that the intersection method would be an alternative to Shor's algorithm for solving EC-DLP \cite{S99} in the post-quantum paradigm. Nonetheless, we propose the intersection method as an alternative to the 'summation polynomial' attack. The reason for such a tentative claim is straight-forward and it stems from the fact that there are no quantum algorithms for computing summation polynomials in an efficient way.  

\newpage

\section{Intersection of Curve Method}

\noindent Let, $E({\mathbb{F}}_{q})$ be an Elliptic Curve defined over a finite field of order $q$ where $q$ is a prime power, for some prime $p$. $P, Q \ \in \ E ({\mathbb{F}}_{q})$ and $n \ \in \ \mathbb{Z}/N\mathbb{Z}$ and $N = ord (P)$. We can define the EC-DLP in this setting as:

	$$Q = nP$$

\noindent, where $n \ < \ \# E({\mathbb{F}}_{q})$	By Hasse's Theorem we have that  $\# E({\mathbb{F}}_{q}) \sim \ q$. The order of $P$ is  known, since it can be quite efficiently computed by the SEA algorithm. Let, $m = \lfloor {\log}_{2} (N) \rfloor$. Now, we can express $n$ as:

	$$n = \sum_{i = 0}^{m} {\epsilon}_{i} {2}^{i} \hspace{20mm} {\epsilon}_{i} \ \in \ \{0,1\}$$
	
\noindent Then, we define $P_i$ as:

	$$P_i := 2^i P \hspace{10mm} \forall i = 0, ... ,m$$	
	
\noindent It follows that:

    $$Q = \sum_{i=0}^{m} {\epsilon}_{i} P_i$$
	
\noindent Let $K(E)$ be the set of all rational functions. Note that if we are able to find a function $f  \in  K(E)$ such that the support of $\div (f)$ is contained in $\{P_0 , ... , P_m , -Q , \O \}$, then we are able to find a relation among $P$ and $Q$. This is the idea of our geometric approach for solving the DLP. To illustrate this fact we provide a well known theorem from basic algebraic geometry:

\vspace{2mm}

\noindent   \textbf{Theorem A} \cite{SR94}: Let $X$ be an affine variety. Then a rational function which is regular at all points of $X$ can be described as a polynomial function.  

\vspace{2mm}

\noindent \textbf{Proposition}: Let, $C: f(x,y) = 0$ be an affine curve and $r(x):= {\text{Res}}_{y} (f(x,y),y^2 -(x^3 + Ax +B))$ and $P_0 = ( x_0 , y_0 ) \ \in \ E$. Then:

    $$v_{(x - x_0)} (r(x)) = {(C \cdot E)}_{P_0} + {(C \cdot E)}_{-P_0}$$

\noindent   \textbf{Proof}: It is a 2-step proof:

\vspace{2mm}

\noindent   \textit{Step 1}: To reduce the proposition to the case where there are no points of $C \cap E$ on the same vertical line $(x = x_i)$

\vspace{2mm}

\noindent   \textbf(Step 2): Now we prove that:

    $${v}_{x - x_0} (r(x)) = {(C \cdot E )}_{P_0}$$
    
\newpage    

\noindent   \textit{Proof of Step 1:}

\vspace{2mm}

\noindent Let, $P_0$ and $- P_0 \ \in \ C \cup E$

\vspace{2mm}

\noindent Then considering $\dfrac{f(x,y)}{x - x_0}$ is regular on the affine part of $E$. Therefore, $\exists$ a polynomial $g(x,y)$ such that $g(x,y) = \dfrac{f(x,y)}{x - x_0}$ on $E ({\overline{\mathbb { F}}}_{q})$ or equivalently $f(x,y) = (x - x_0)g(x,y) \ \ (mod \ E)$ [Invoking Theorem A]. The resultant is multiplicative and from the above argument, it follows that:

    $${\Res}_{y} (E , f(x,y)) = {\Res}_{y} (E , x - x_0) {\Res}_{y} (E , g(x,y))$$
    
    $${\Res}_{y} (E , f(x,y)) = {(x - x_0)}^{2} {\Res}_{y} (E , g(x,y))$$

\noindent Iterating the process we get the reduction which proves our assertion for the first step.

\vspace{2mm}

\noindent   \textit{Proof of Step 2:}

\vspace{2mm}

\noindent Now, we can safely assume that there are no points of $C \cap E$ in the same vertical line. Hence, the assertion becomes:

    $${v}_{x - x_0} (r(x)) = {(C \cdot E )}_{P_0}$$

\noindent We observe that if all the points are simple then the 'Resultant' is a square-free polynomial which is given by:

    $$\prod_{P \in C \cap E} (x - x_P)$$ 
    
\noindent   (Where all the $x_p$ are different)

\vspace{2mm}

\noindent Let's consider the case where we have a tangent $({T}_{P_0})$ at $P_0$ then:

    $${v}_{x - x_0} = ({\Res}_{y} (E, {T}_{P_0})) = 2 \ or \ 3$$ 
    
\noindent (Depends if the point is a flex or not!)

\vspace{2mm}

\noindent We prove the step $2$ by induction on:

    $$M = \sum_{P \in C \cap E} {(C \cdot E)}_{P} - 1$$

\noindent If, $M = 0$ then the points are simple. Let's suppose that  $M > 0$, and let $P_0$ be a point such that the ${(C \cdot E)}_{P_0} \ \geq \ 2$. Then, considering $\dfrac{f(x,y)}{{T}_{P_0}}$ we have $3$ possibilities:

\noindent 1) Considering $P_0$ is not a flex. If, all the points of intersection belong to $C \cap E$ and $P_0$ is not a flex. Then, $\dfrac{f(x,y)}{{T}_{P_0}}$ is a regular function. Hence, by the previous argument we can reduce $M$ and apply induction accordingly to deduce the result.

\vspace{2mm}

\noindent 2) If $P_0$ is not a flex but we have a third point of intersection $(R) \ \notin \ C \cap E$. Then, considering a line $L_R$ passing through $R$ and through other two points whose $x$-coordinates are different from each other and also different from that of the points in $C \cap E$. Now, we deduce that $\left(\dfrac{f(x,y)}{{T}_{P_0}}\right) ( L_R )$ is regular and the value of $M$ for the  divisor associated to this function has reduced.

\vspace{2mm}

\noindent   3) If $P_0$ is a flex. Then, we have $2$ sub-cases:

\vspace{2mm}

    3a) \hspace{40mm} ${(C \cdot E)}_{P_0} \ \geq \ 3$

\vspace{2mm}

\noindent   In this case, it is a straightforward deduction that $\dfrac{f(x,y)}{{T}_{P_0}}$ is regular in the affine part of $E$ and in this way we have reduced the value of $M$ and apply induction accordingly to deduce the result.

\vspace{2mm}

    3b) \hspace{40mm} ${(C \cdot E)}_{P_0} = 2$
    
\vspace{2mm}    
    
\noindent   For this case we consider a line passing through $P_0$ and other two points whose $x$-coordinates are different from each other and also different from that of the points in $C \cap E$. Now, we deduce that $\left( \dfrac{f(x,y)}{{T}_{P_0}} \right) ({L}_{P_0})$ is regular and the value of $M$ for the divisor associated to this function has reduced.

\vspace{2mm}

\noindent   This ends the proof of the proposition. We continue with the explanation of the intersection method. 

\vspace{2mm}

\noindent Suppose that the solution $n$ to the DLP is even and let:
    
    $$\frac{n}{2} = \sum_{i = 0}^{m - 1} {\varepsilon}_{i} \cdot 2^i$$
    
\noindent This is the base 2 decomposition. Note that this is not a huge restriction as the algorithm can check for $P = (n - 1)Q$ in case we do not find an even $n$. Now, we recall the group isomorphim:

	$$E \rightarrow {Pic}^{0} (E)$$

    $$P \ \mapsto \ (P) - (\mathcal{O})$$

\noindent Here, we want to highlight that in ${Pic}^{0} (E)$ we can view the EC-DLP in terms of divisors. The idea here is to use the concept of principal divisors which enables us to write the relations of the following form:

    $$\sum_{i} {P}_{i} = \O$$

\noindent   Hence, we can be sure that we can write these relationships in terms of the existence of polynomial functions $f(x,y)$ on $E$ such that the $\div (f) = \sum_{i} ({P}_{i}) - (\O)$ in the group of divisors which implies that the restriction to the affine part of $E$ is equal to $\sum_{i} ({P}_{i})$. Since, $n = \sum_{i=0}^{m-1} 2 {\varepsilon}_{i} \cdot 2^i$. We have that there exists a function $f(x,y) \in K(E)$ and regular in the affine part of $E$, such that:

    $$\div(f) = (-2m - 1) (\O) + (-Q) + \sum_{i=0}^{m-1} (1 + {\varepsilon}_{i}) ( P_i ) + ( 1 - {\varepsilon}_{i}) ( - P_i )$$

\vspace{2mm}

\noindent Note that, by the previous proposition we have that:

    $${\Res}_{y} (f(x,y),y^2 - (x^3 + Ax + B)) = {( x - x_0 )}^{2} \cdots {(x - {x}_{m - 1})}^{2} ( x - x_Q )$$

\noindent Where, $P_i = ( x_i , y_i )$, $Q=( x_Q , -y_Q )$.  Since,  $f$ is regular in the affine part of $E$, it is an element of ${\mathbb{F}}_{q}[x,y]/(y^2 - (x^3 + Ax + B))$, which is uniquely determined by a polynomial of the form:

    $$f(x,y) = yg(x) + h(x)$$
    
\noindent   Let, $d = \text{deg}(f(x,y)) = d$, we can write:
    
    $$g(x) = g_0 + ... + {g}_{d - 1} {x}^{d - 1}$$
    
    $$h(x) = h_0 + ... + h_d x^d$$

\noindent   Hence, the functions which are regular on the affine part can be parametrized up to constant by a projective space ${\P}^{2d}$. The resultant can be expressed (up to multiplication by a non-zero constant) in the following way:

    $$r(x)= {(h(x))}^{2} - {(g(x))}^{2} (x^3 + Ax + B)$$
    
\noindent So, if ${g}_{d - 1} \ \neq \ 0$ and $r(x)$ is an uni-variate polynomial of degree $2d + 1$. It follows that imposing $d = m$, ${g}_{d - 1} = 1$ and that:

    $$r(x) = {( x -x_0 )}^{2} \cdots {(x - {x}_{m - 1})}^{2} ( x - x_Q )$$

\noindent   We are imposing the condition that, $2m + 1 = 2d + 1$ algebraic conditions (more precisely quadratic conditions) on an affine space of dimension $2m$. Let, $I \ \triangleleft \ \kappa [ g_0 , ... , {g}_{d - 2}, h_0 , ... , h_d ]$ be the ideal generated by these equations. This results in an over-determined system which has no solution, in general. This is not surprising, since in general there is no solution for the EC-DLP. The Diffie-Helmann Key exchange which is the basic example of posing EC-DLP would not be possible if there is no solution. Hence, it can be safely assumed that for such a system for specific/recommended curves will have solution. But if there is a solution, then they are at least 2, in fact:

    $$( g_0 , ... , {g}_{d - 2}, h_0 , ... , h_d ) \ \in \ V(I) \ \iff \ ( g_0 , ... , {g}_{d - 2}, -h_0 , ... , -h_d ) \ \in \ V(I)$$

\noindent This corresponds to the fact that if:
    
    $$Q= \sum_{i = 0}^{m - 1} [2 {\varepsilon}_{i}] P_i \ \iff \ -Q = \sum_{i = 0}^{m - 1} [-2 {\varepsilon}_{i}] P_i$$

\noindent In general, there can be also more solutions if we do not require that $f(P_i) = 0$ and $f(-Q) = 0$, which are $m + 1 = d + 1$ linear conditions on the coefficients of $f$. So, at the end we have a variety in ${{\A}^{2d}}_{\overline{\F_q}}$ defined by $d + 1$ linear equations and $2d + 1$ quadratic equations. Note that some of the quadratic conditions are redundant. One of the instances is the condition $f( x_i , y_i ) = 0$ and $r( x_Q ) = 0$ which imply that $r(x_i) = 0$ and $r(x_Q) = 0$. Therefore, there are $d + 1$ linear conditions and $d$ independent quadratic conditions given by:
    
    $$r'(x_i) = 0, \hspace{10mm} \forall \ i \ \in \ \{0,...,m-1\}$$

\noindent   Thus, essentially we are studying the zero set of the interesection of $d$ quadric hypersurfaces in an affine space of dimension $d - 1$.

\vspace{2mm}

\noindent   \textbf{Remark}: Note that if $Order(P)$ is even and $n$ is odd this algorithm produces \textit{no solution}. Instead if $n$ is even then there can be two solutions if $n + Order(P)$ has the same binary length of $Order(P)$. If $Order(P)$ is odd (it is very common that $\# E(\mathbb{F}_p))$ is chosen to be a prime number for security reasons) then there is a unique solution provided $n$ is even. If $n$ is $odd$ there is exactly 1 solution corresponding to $n + Order(P)$ if this number has the same binary length of $Order(P)$, no solutions otherwise.

\section{Algorithm}

\noindent   The algorithm consists of steps which depends on accessing the points on the elliptic curve and operating on them to implement the idea of intersection of surfaces. It starts with initialization of the finite field and the elliptic curve defined over it. Then we generate a polynomial ring and define the required polynomials. The vanishing ideals are evaluated henceforth and the derivative is computed. After this Grobner basis is computed and the F4 algorithm is implemented to solve the system of equations. The algorithm is realized through MAGMA. One can find the MAGMA code \href{https://www.dropbox.com/s/h56m7ah6wwfp9n2/Intersection%20Method.txt?dl=0}{here}. We would like to emphasize that we have not implemented '\textit{any}' quantum algorithm. The implementation is purely classical. 
We hereby provide the pseudocode for the the sub-routines of the algorithm. 

\newpage

\begin{algorithm}
\begingroup
\footnotesize
	
	\SetAlgoLined
	
	construct FiniteField(p,k), EllipticCurve([F | A,B]) and $P \ \in \ E$  \tcc{p is a prime number and A, B are the coefficients of the elliptic curve equation} 
	
	\For{$s = 0;\ s < m_1 ;\ s = s + 1$}{define ps as [$2^s * P$]}    \tcc{Operation on the point P. $m = Floor(Log (2, n)), m_1 = m - 1, m_2 = m - 2$ and similarly we can continue} 
	
	construct PolynomialRing(F, $2*m$) (R) and PolynomialRing(R, 2) \tcc{The variable is w and k respetively}
	
	\For{$i = 1;\ i < m;\ i = i + 1$}{define h as w[i]} \tcc{Array of coefficients}
	
	construct Polynomial(h) (H) and Polynomial(g) (G)   \tcc{Variables are u and v respectively}
	
	Evaluate(H, k[1]) and Evaluate(G, k[1]) + $k[1]^( m_2 )$    \tcc{Evaluating the Polynomial}
	
	\For{$i = 1; \ i < (\#T + 1); \ i = i + 1$}{define Z as [ Evaluate(f, T[i]) ]}  \tcc{Array ps is called and an array of arrays is constructed and named T. f is defined as the addition of the evaluation in previous steps}
	
	construct an ideal $i_1$ as < R | Z >  \tcc{Initializing the ideal for calculating the vanishing ideal. $i_2 , i_3 , i_4 , i_5$ ideals are constructed consequently} 
	
	Evaluate Derivative(r, k[1])    \tcc{r is $x1^2 - k[1]^3 * y1^2 - A*k[1]*y1^2 - B*y1^2$}
	
	initialize Q as $(n - 100)*P$	\\ 
	
	compute Radical( $i_5$ ), Groebner(R), PrimaryDecomposition(R), GroebnerBasis(R[1]) \\

	\For{$i = 1; \ i < (\# Z + 1); \ i = i + 1$}{
		\If{Z[i] = 0}{Set $Z_1$ = [0]}
		\Else{set $Z_1$ = [ Z[i] ]}}   \tcc{Z is constructed as the array of arrays of the primary decomposition}
	
	\For{$i = 1; \ i < (\# Z + 1); \ i = i + 1$}{
		\If{$Z_1$[i] = 0}{set $Z_2$ = [0]}
		\Else{set $Z_2$ = [ $2^{( Z_1 [i])}$]}}
	
\caption{Intersection Method}

\endgroup

\end{algorithm}

\section{Conclusion and Future Work}

\noindent   The complexity of the curve intersection algorithm depends majorly on finding the vanishing ideals and the system of equations which are over-determined in nature. Hence, using classical tools and technique, the complexity is quite high. As we have highlighted that the proposed method produces overdetermined system of equations which are hard to solve. In the classical paradigm, machine learning technique gives us hope to solve these kind of system with a comparatively lower (probabilistic) time complexity. Another approach which is worth exploring is Zhuang-Zi \cite{DGS06} which provides a new technique to solve system of multivariate polynomial equations over a finite field. In the post quantum paradigm, efficient methods for solving these system of equations are illustrated in \cite{FHKKKP17} (only for boolean case), and \cite{HHL09} (linear system of equations). \cite{AD07} entails a quantum algorithm for solving non linear system of equations in $GF( q = p^k )$ which might be used for improving the proposed method in the future. We reiterate that \textit{we do not expect that the intersection method would be an alternative to Shor's algorithm for solving EC-DLP} \cite{S99} in the post-quantum paradigm. Nonetheless, we propose the intersection method as an alternative to the 'summation polynomial' attack. The reason for such a tentative claim is straight-forward and it stems from the heuristics that the proposed method is fast in reducing EC-DLP into a over-determined system of equations which we expect to be solved efficiently by a quantum computer.

\section{Acknowledgements}

\noindent We would like to thank Professor Gerhard Frey for patiently hearing about our idea during his visit to Roma Tre University. He encouraged us to pursue this path and we are grateful to him for his time and attention. We would like to thank the High Performance Computing (HPC) facility of University of L'Aquila, which enabled us to implement our algorithm on MAGMA and run the experiments to validate our results. We are grateful to the fruitful and illuminating discussions with Professor Norberto Gavioli of University of L'Aquila.

\end{document}